# Study of the pressure effects in TiOCl by ab initio calculations

A. Piñeiro[a,b,*], V. Pardo[a,b], D. Baldomir[a,b], S. Blanco-Canosa[c], F. Rivadulla[c], J. E. Arias[b], J. Rivas[a]

[a]*Departamento de Física Aplicada, Universidade de Santiago de Compostela, E-15782, Santiago de Compostela, Spain*
[b]*Instituto de Investigacións Tecnolóxicas, Universidade de Santiago de Compostela, E-15782, Santiago de Compostela, Spain*
[c]*Departamento de Química-Física, Universidade de Santiago de Compostela, E-15782, Santiago de Compostela, Spain*

A B S T R A C T

Electronic structure calculations on the low dimensional spin-1/2 compound TiOCl were performed at several pressures in the orthorhombic phase, finding that the structure is quasi-one-dimensional. The $Ti^{3+}$ ($d^1$) ions have one $t_{2g}$ orbital occupied ($d_{yz}$) with a large hopping integral along the b direction of the crystal. The most important magnetic coupling is Ti-Ti along the b axis. The transition temperature ($T_c$) has a linear evolution with pressure, and at about 10 GPa this $T_c$ is close to room temperature, leading to a room temperature spin-Peierls insulator-insulator transition, with an important reduction of the charge gap in agreement with the experiment. On the high-pressure monoclinic phase, TiOCl presents two possible dimerized structures, with a long or short dimerization. Long dimerized state occurs above 15 GPa, and below this pressure the short dimerized structure is the more stable phase.

2009.

## 1. Introduction

In recent years, much attention has been drawn to the oxyhalides TiOCl and TiOBr, specially after the finding of these materials undergoing a spin-Peierls transition at low temperature [1] brought about by the strong one-dimensionality of the material. The low-dimensional spin-1/2 compound TiOCl shows two consecutive phase transitions at the incommensurate spin-Peierls temperature ($T_{ISP}$~91 K) and at the commensurate spin-Peierls temperature ($T_{SP}$~66 K) [2][3]. A coupling of a one dimensional antiferromagnetic S=1/2 chain with the lattice results in a spin-Peierls transition with a nonmagnetic (singlet) dimerized ground state. $CuGeO_3$ is the first well established example for such a transition in an inorganic compound [4][5][6].

On the other hand, Kuntscher *et al* [7] observed a strong suppression of the transmittance and an abrupt increase of the near-infrared reflectance above ~10 GPa in TiOCl. These effects were interpreted as a pressure-induced metallization of the 1D chain. Recently, the strong suppression of the electronic gap above ~12 GPa was confirmed directly through high-pressure resistivity [8], although the semiconducting behaviour at high pressures is still maintained ($E_g$~0.3 eV). On the other hand, Blanco-Canosa *et al*. [9] found evidences of a spin-Peierls to Peierls transition as a function of pressure, due to an enhanced dimerization of the Ti chain along the b-axis that puts the material close to the itinerant electron limit. This effect was observed previously in other similar materials near a metal-insulator transition [10][11][12].

In this work we study from ab initio calculations the evolution with pressure of the chemical and electronic structure of TiOCl. It is very important to corroborate the quasi-1D electronic nature and to study the evolution of the transition temperature with pressure. We will show here that the structural transition from the low-pressure orthorhombic to the high-pressure monoclinic dimerized phase is probably more complex than anticipated due to the possible existence of two dimerized phases on the monoclinic structure with different short and long bonds between Ti atoms on the b-direction chains.

## 2. Structure

TiOCl crystallizes in an orthorhombic quasi two dimensional structure (FeOCl-type) where buckled Ti-O bilayers within the ab plane are well separated by $Cl^-$ ions (see Fig. 1(a)). Ti atoms form one-dimensional chains along b direction, and the planes pile up along the c axis. Each Ti atom is coordinated by a heavily distorted octahedron formed by 4 O and 2 Cl atoms. Magnetic neighbours are 2 Ti along b (large hopping), 2 Ti along a (far enough for neglecting that magnetic interaction) and 4 closest Ti atoms out of the ab plane. Lattice parameters were measured using high pressure X-ray diffraction for pressures up to 9 GPa, in the orthorhombic space group Pmmn [9]. Using the experimental lattice parameters, we optimized the atomic positions ab initio for each of the pressures considered

This system presents a structural transition to a monoclinic phase al low temperature [13], which is associated to the spin-Peierls transition [2]. High pressure X-ray diffraction at 15 GPa confirms that this monoclinic structure (Fig. 1(b)) also occurs at high pressure at ambient temperature [9]. We have also studied this monoclinic phase at intermediate pressures (this phase is stable below room temperature for P< 15 GPa). For carrying out the corresponding ab initio calculations at the pressures presented, we have made an extrapolation of the lattice parameters in the space group $P2_1/m$ based on a study of its evolution of Forthaus *et al*. [8]. We start with experimental input for the lattice parameters at ambient pressure and low temperature, and at 15 GPa and room temperature.

_______________
* Corresponding author. Tel.: +34-981563100 (13960); fax: +34-981520676.
*E-mail address*: alberto.pineiro@usc.es.

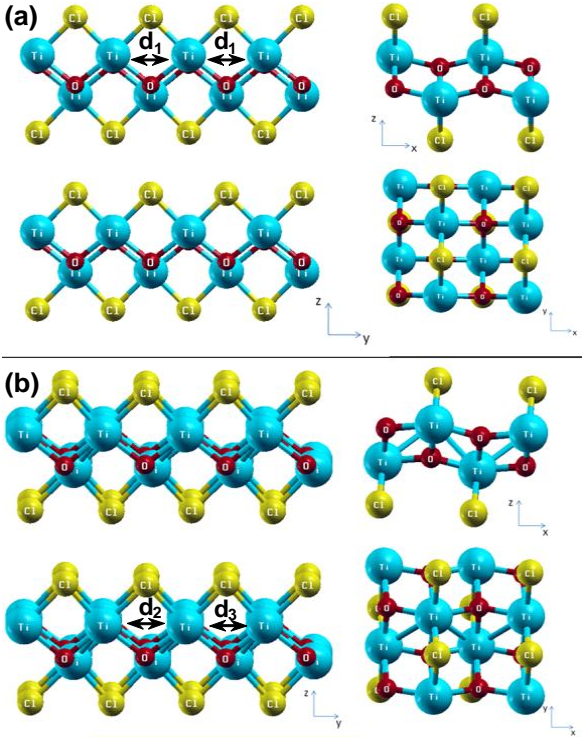

**Fig. 1**. Structure of the TiOCl compound (a) on the ambient pressure orthorhombic phase and (b) on the high-pressure monoclinic phase. Values of the distances are $d_1$=3.36Å, $d_2$=3.25Å and $d_3$=3.39Å.

## 3. Computational details

We employ here full potential, all electron, electronic structure calculations based on the density functional theory utilizing the APW+lo method [14] performed using the WIEN2k software [15][16].

For modeling the behavior of the d electrons of the system, we included the strong correlation effects by means of the LDA+U scheme [17], where the correlation effects are controlled by an effective U ($U_{eff}$=U-J), being U the on-site Coulomb repulsion and J the on-site exchange constant (taken as J= 0, as is common practice in literature). A value of $U_{eff}$ = 5 eV was used in the calculations since it reproduces the ambient pressure band gap, but the results presented are consistent for values of $U_{eff}$ from 4 to 7 eV. The structural minimization was carried out using the GGA-PBE scheme [18]. The parameters of our calculation were fully converged for every particular case to the required precision.

## 4. Results

Figure 2 shows the electron spin density of TiOCl in the orthorhombic phase. Since basically all spin moment is due to the Ti ions, this plot helps us to understand the electronic structure in the vicinity of the Fermi level. We can see the electronic structure of TiOCl and its quasi-one-dimensional nature. The $Ti^{3+}$ ($d^1$) ions have one $t_{2g}$ orbital occupied ($d_{yz}$) with a large hopping integral along the b direction of the crystal, leading to a highly one-dimensional electronic structure. The interactions along the a-axis are very small due to the symmetry of the occupied orbital: along the c axis, interactions are negligible due to the large Ti-Ti distance. Hence, there is an electronic reduction of the dimensionality of the system, which eventually would explain the spin-Peierls type distortion the system undergoes.

We have also calculated the different magnetic couplings in the structure and their variations with pressure. Our results are summarized in Table I. To obtain the different exchange constants, we have calculated the total energies within the LDA+U scheme for various magnetic configurations and several pressures within the orthorhombic structure (P<9 GPa). Due to the structure of the material, we consider three different magnetic couplings: $J_d$, within the one-dimensional chains along the b axis; $J_s$, the coupling between a Ti and its 4 closest neighbors and $J_l$, the coupling along the a axis. We have fit our total energies to a Heisenberg-type Hamiltonian of the type H=$\sum J_{i,j} S_i S_j$ (where the sum runs over spin pairs). With such a Hamiltonian, J positive means antiferromagnetic coupling is favored and J negative means ferromagnetic coupling is favoured.

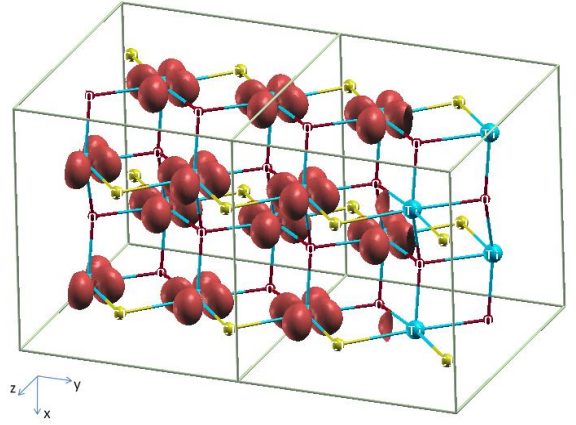

**Fig. 2**. Electron spin density plot of TiOCl in the orthorhombic structure. x, y and z axis corresponds to a, b and c directions respectively.

We observed that the coupling along the b axis is much bigger than the others ([$J_b/J_a$]~300 and [$J_b/J_s$]~25) due to the highly-1D electronic structure of the material and increases rapidly with pressure. The calculated in-chain coupling is in good agreement with fittings to a spin-1/2 Heisenberg chain model of the experimental susceptibility curves[19][20][21] (view Table I). This coupling will give rise to the low-temperature spin-Peierls phase. The increase with pressure of the magnetic exchange constant along the b axis indicates a large increase with pressure of the associated transition temperatures to the spin-Peierls phase, and more important an increment of the one-dimensionality of the system with pressure. Fitting to a 1D-Heisenberg model ($K_B T_c$=(1/3)z$JS^2$, where z is the number of nearest neighbours) yields values of the transition temperatures close to those obtained experimentally[19], noting that the exchange coupling constants we calculate are one half of those due to the different choice of the Hamiltonian we made. For P>10 GPa, a transition is expected at room temperature, as has been found experimentally [9]. Our theoretically estimated transition temperature of about 275 K for P= 9 GPa also agrees with experiment. Hence, our calculations explain how the evolution with pressure of this magnetic coupling is associated to the transition temperature to a spin-Peierls phase. The appearance of a dimerized spin-Peierls phase at room temperature when pressure is increased above 10 GPa is a consequence of the large pressure dependence of the magnetic exchange coupling along the b axis, due to the large one-dimensionality of the electronic structure and the small Ti-Ti distance along the b axis and the strong σ-type bond along that direction.

**Table I**. Magnetic couplings between Ti atoms and theoretical evolution of $T_c$ with pressure. These $T_c$ values were calculated fitting to a 1D-Heisenberg model $K_B T_c$=(1/3)z$JS^2$, where z is the number of nearest neighbours. J positive means antiferromagnetic coupling and J negative means ferromagnetic coupling. Pressure values are in GPa, magnetic couplings and $T_c$ are in Kelvin.

| P (GPa) | 0 | 4 | 7 | 9 |
|---|---|---|---|---|
| $J_s$(K) | -12 | -10 | 0 | 10 |
| $J_d$(K) | 300 | 1100 | 1400 | 1650 |
| $J_l$(K) | -1 | -4 | -4 | -4 |
| $T_c$(K) | 50 | 180 | 230 | 275 |

Two types of dimerized structures could be converged for this material in a monoclinic space group. At P= 0, the Ti-Ti distances along the b axis are 3.25Å in the short bond and 3.43Å in the long bond, a difference about 3%, but at higher pressures (not stable at P= 0GPa), a different type of dimerization can be analyzed, which we have named "long-dimerized". In this structure Ti-Ti distances along the b axis is about 2.95Å in the short bond and is about 3.69Å in the long bond, a difference about 11% (values calculated at P= 15 GPa, where this structure becomes more stable). Hence, at every pressure, we have compared the total energy of these two possible dimerized structures. Results are summarized in Fig. 3. Above P=15 GPa a transition occurs to a high pressure "long dimerized" phase. At a similar pressure, an anomaly is found in the resistivity measurements [8], which can be explained by a strong reduction of the gap in such a structure [9].

The discontinuity in the evolution of the bond-distances in the dimerized phase at high pressures is at odds with expectations in a conventional spin-Peierls scenario. On the other hand, as the short Ti-Ti distance approaches the limit for electron-itinerancy, the spin-Peierls distortion is expected to be supported by a conventional Peierls distortion of the 1D chain, hence increasing the difference between the short and long bonds. So, it is most probably the proximity of the material to the itinerant electron limit at high pressure what drives the transition from the short-to-long dimerized structure.

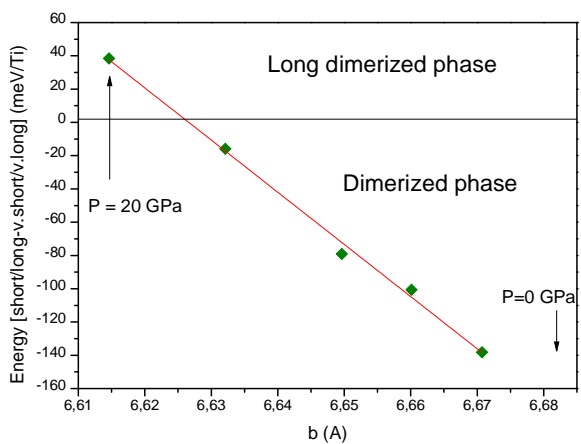

**Fig. 3**. Difference between total energy of "short" dimerized structure and "long" dimerized structure at several pressures. We can see the transition at about P=15 GPa.

## 5. Summary

We have carried out ab initio calculations on TiOCl, based on experimental data of the evolution of the structure with pressure. These experiments evidence the existence of a structural transition at high pressure on TiOCl. We have characterized the structural and physical properties of the high pressure phase, where $Ti^{3+}$-$Ti^{3+}$ dimerization is stronger (i.e., there is a much larger difference between the large and short Ti-Ti distance) than in the ambient pressure phase. We have analyzed the quasi-one-dimensional electronic structure of TiOCl and also the strong one-dimensionality of the magnetic properties. $Ti^{3+}$ ($d^1$) ions have one $t_{2g}$ orbital occupied ($d_{yz}$) with a large hopping integral along the b direction of the crystal. Our calculations predict the transition to a high-pressure dimerized phase at room temperature at abut 10 GPa, in accordance with the experiment. At around 15 GPa, the short Ti-Ti distance approaches the critical distances for metallic bonding, and the material undergoes a transition from a spin-Peierls phase to a more conventional Peierls transition.

**Acknowledgments**

The authors thank the CESGA (Centro de Supercomputación de Galicia) for the computing facilities, the Ministerio de Educación y Ciencia (MEC) and Xunta de Galicia for finacial support through the projects MAT2006/10027, HA2006-0119 and PXIB20919PR respectively. F.R. and S. B-C. also thank MEC of Spain for support under Ramón y Cajal and FPU programs respectively. A.P. thank the Universidade de Santiago de Compostela for financial support.